\documentclass[twoside,a4paper,11pt]{sca}
\usepackage{graphicx}
\usepackage{hyperref}
\usepackage{natbib}  
\begin{document}
\pagenumbering{arabic}
\pagestyle{myheadings}
\thispagestyle{empty}
{\flushright\includegraphics[width=\textwidth,bb=90 650 520 700]{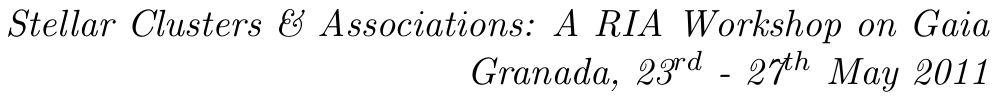}}
\vspace*{0.2cm}
\begin{flushleft}
{\bf {\LARGE
%
Probing the near-IR flux excess in young star clusters
%
}\\
\vspace*{1cm}
%
Adamo Angela$^{1}$,
{\"O}stlin G{\"oran}$^{1}$, 
and 
Zackrisson Erik$^{1}$
%
}\\
\vspace*{0.5cm}
%
$^{1}$
Department of Astronomy, Stockholm University, Oscar Klein Center, AlbaNova, Stockholm SE-106 91, Sweden\\
%
\end{flushleft}
%
\markboth{
NIR excess in young SCs
}{ 
%
Adamo, A. et al.
%
}
\thispagestyle{empty}
\vspace*{0.4cm}
\begin{minipage}[l]{0.09\textwidth}
\ 
\end{minipage}
\begin{minipage}[r]{0.9\textwidth}
\vspace{1cm}
\section*{Abstract}{\small
%
We report the results of a recent study of young star clusters (YSCs) in luminous blue compact galaxies (BCGs). The age distributions of the YSCs suggest that the starburst episode in Haro 11, ESO 185-IG13, and Mrk 930 started not more than 30-40 Myr ago. A peak of cluster formation only 3 - 4 Myr old is observed, unveiling a unique sample of clusters still partially embedded. A considerable fraction of clusters (30 - 50 \%), mainly younger than 10 Myr, shows an observed flux excess between 0.8 and 2.2 $\mu$m. This so-called near-infrared (NIR) excess is impossible to reproduce even with the most recent spectral synthesis models (that include a self-consistent treatment of the photoionized gas). We have used these YSCs to probe the very early evolution phase of star clusters. In all the three host galaxies, the analysis is limited to the optically brightest objects, i.e., systems that are only partially embedded by their natal cocoons (since deeply embedded clusters are probably too faint to be detected). We discuss possible explanations for this NIR excess, in the context of IR studies of both extragalactic young star clusters and resolved massive star forming regions in the Milky Way and in the nearby Magellanic Clouds.

%
\normalsize}
\end{minipage}
%
%
%
\section{Introduction \label{intro}}

Young star clusters (YSCs) are considered a powerful tool to study the star formation history of their hosts.  However, many assumptions and constraints on the evolutionary properties are needed in order to compare YSCs properties and the bulk of the host stellar population. In nearby galaxies, it is possible to study both the resolved stellar population and cluster population \citep[e.g.][]{2011A&A...529A..25S}. In farther galaxies, however, the study of YSCs is conducted with an analysis of their integrated light at different wavelengths under the assumption of  an instantaneous burst, i.e. single stellar population. 

The luminous (M$_B< -17.0$, corresponding to total stellar masses of $\sim 10^{9-10}$ M$_{\odot}$) blue compact galaxies (BCGs) show clear signatures of interactions and/or mergers, and the numerous observed massive YSCs are likely the result of these encounters \citep{2003A&A...408..887O, 2010MNRAS.407..870A, 2011MNRAS.414.1793A, 2011MNRAS.tmp..739A}. The very bright ultraviolet and optical luminosities of these systems suggest rather low dust and metallicity content. Spectra dominated by emission lines clearly demonstrate that BCGs are undergoing a starburst episode. The youth of the burst episode is also observed in the recovered age distribution of the star clusters, which shows a peak of cluster formation younger than 5 Myr. 

The analysis of the young cluster populations in BCGs is quite challenging due to the rapid evolution a cluster experiences during the first 10 Myr (still partly in an embedded phase). Moreover, this analysis is based on the integrated luminosities of the clusters, which appear unresolved at the distance of the targets.  Observations of resolved newly born star clusters in the Milky Way and in the Magellanic Clouds reveal that these are quite complex systems. A cluster  forms in a hierarchical medium, from the fragmentation and collapse of giant molecular cloud complexes \citep{2010IAUS..266....3E}. This implies that stars form in a fractal distribution and only in some cloud cores are their number densities is high enough to eventually form a bound cluster \citep{2010MNRAS.409L..54B}. This picture is confirmed also by dynamical studies. \citet{2011MNRAS.410L...6G} observed a continuous distribution between associations (unbound) and clusters (bound) at very young ages, and a clear distinction at older stages, when the crossing time of unbound systems exceeds their stellar age.

In a newly born cluster, the massive and short-lived stars, rapidly reach the main sequence and produce strong winds and UV radiation, which ionizes the intracluster gas and create bubbles and shells. These H{\sc ii} regions surround the optically bright core of stars and significantly contribute to the integrated fluxes. However, a large fraction of the stars is still accreting material from their dusty disks (young stellar objects, YSOs) or contracting (in the pre-main sequence phase, PMS). Moreover, the edges of the clusters are places for triggered \citep{2011arXiv1101.3112E} and progressive star formation \citep[e.g.][]{2007ApJ...665L.109C}, which can explain the observed age spread in the  star forming regions \citep[e.g.][]{2010ApJ...713..883B}.

\section{Evidence for the NIR excess}

\begin{figure}
\center
\includegraphics[width=7.6cm]{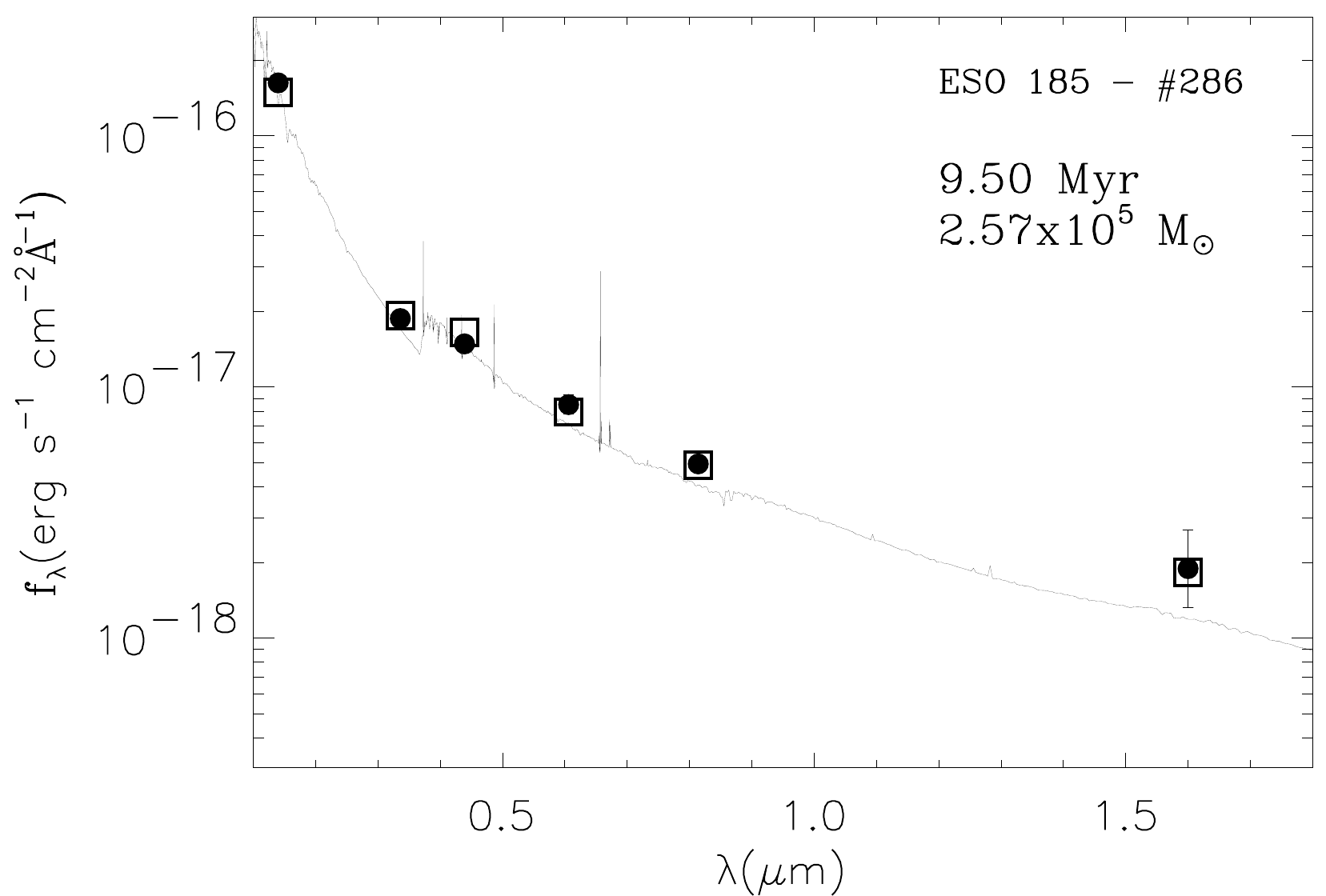} 
\includegraphics[width=7.6cm]{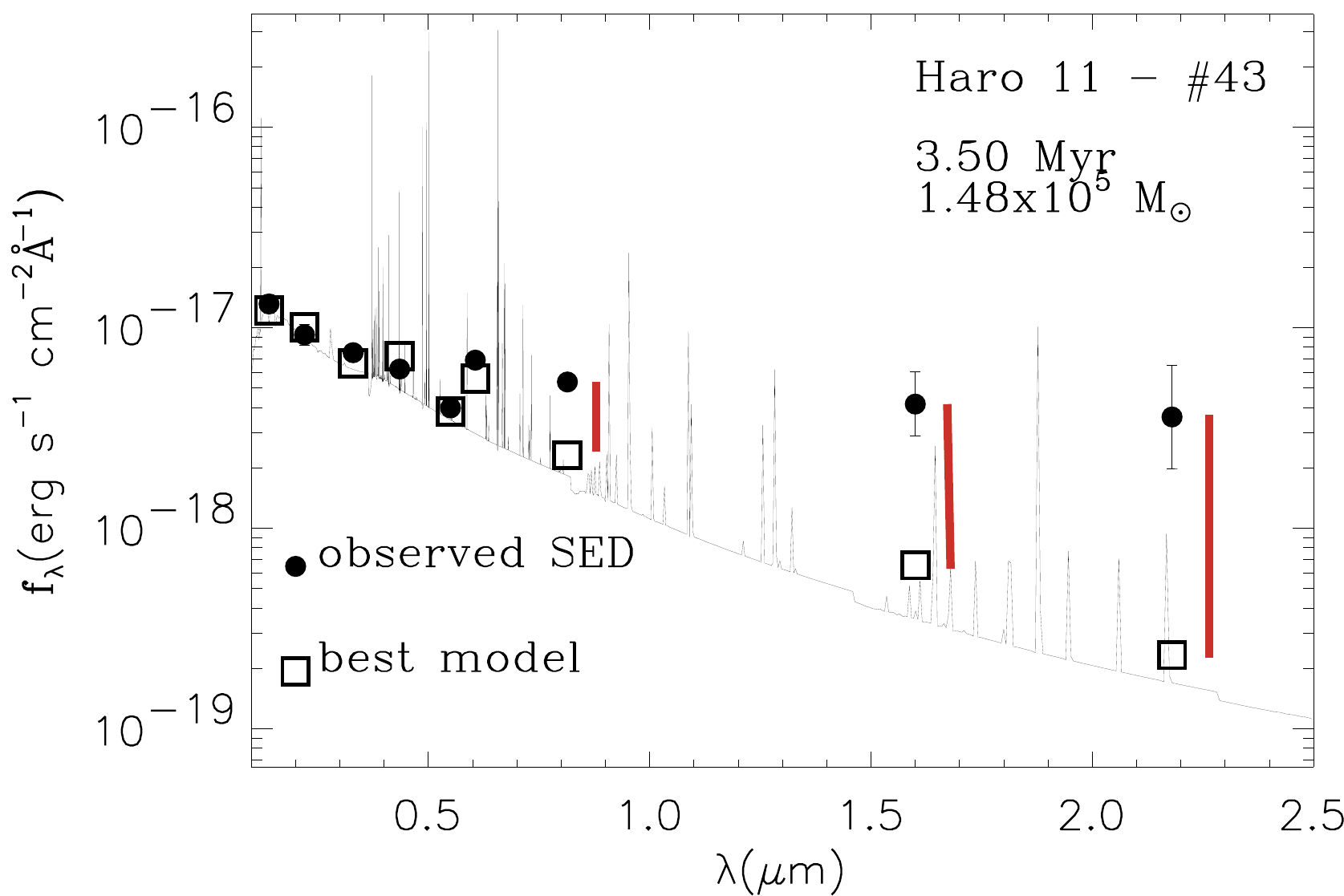} 
\caption{\label{fig1} An example of SED analysis of 2 clusters (the name of the host in indicated in the inset). The two objects are representative of different cases: on the left a young cluster with a good fit, i.e. regular SED; on the right, a cluster affected by an excess in flux at wavelengths longer than 8000 \AA. The filled black points indicated the observed fluxes for each cluster. The integrated model
fluxes are labelled with open squares. They sit above the best fitting spectrum overplotted with a solid line. Red vertical lines indicate which data points have been excluded from the fit. The age and mass of the cluster are shown.}
\end{figure}

\begin{figure}
\center
\includegraphics[width=7.6cm]{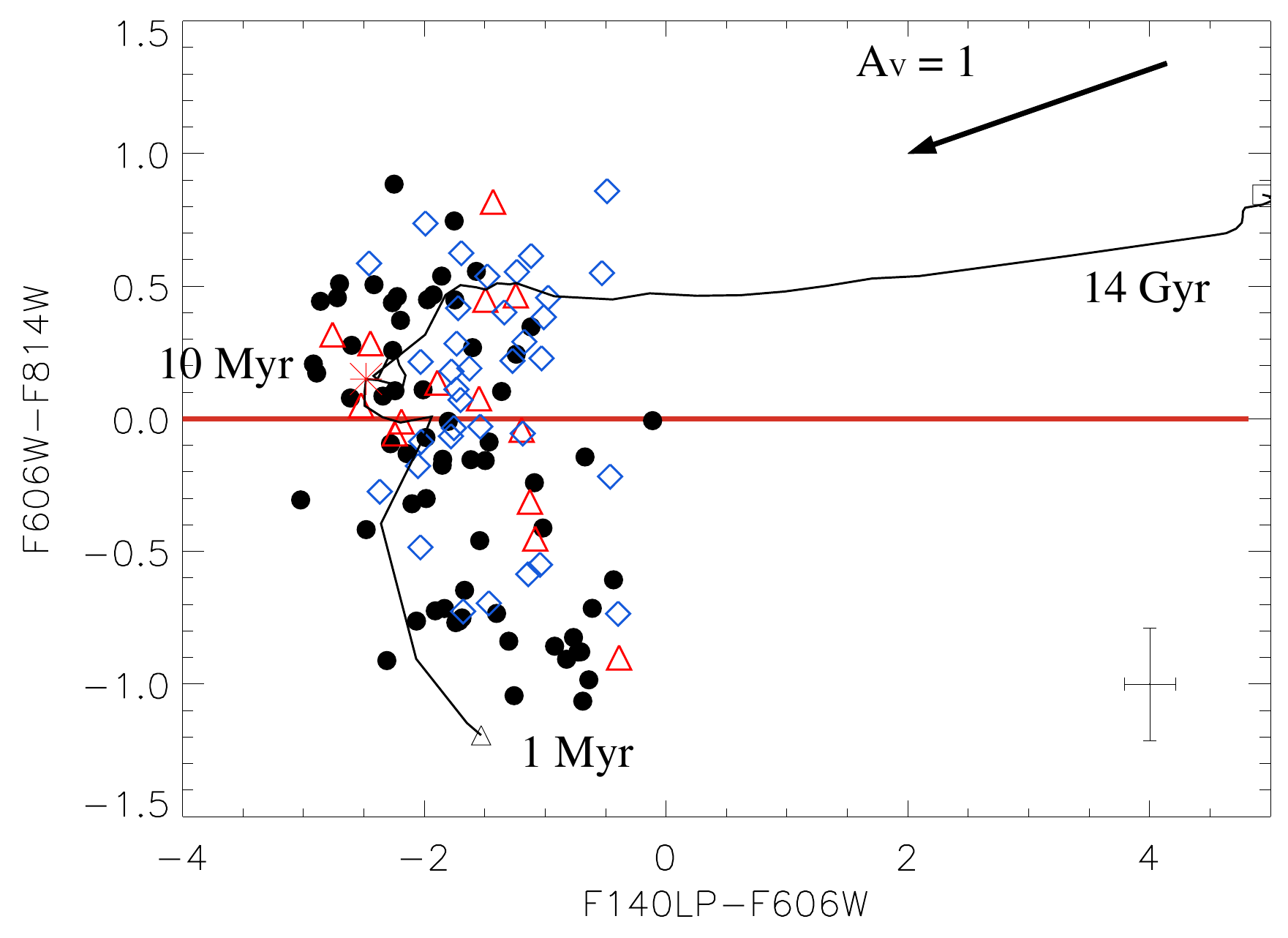} 
\includegraphics[width=7.6cm]{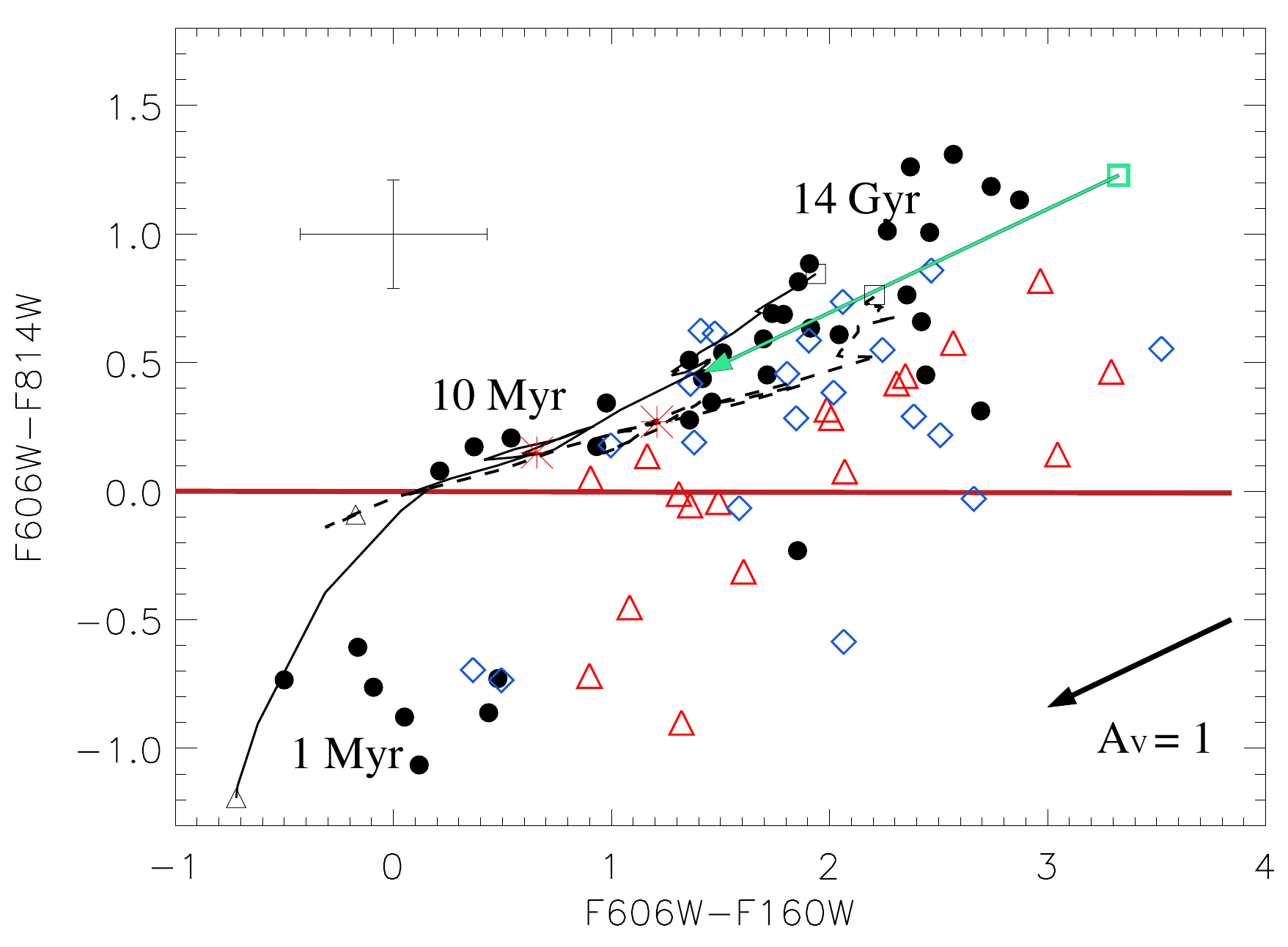} 
\caption{\label{fig2} Color-color diagrams of the cluster population in Mrk\,930 \citep{2011MNRAS.tmp..739A}. Different filter combinations are compared to the $R-I$ color (F606W-F814W). Left: $UV-R$ (F140LP-F606W); Right: $R-H$ (F606W-F160W). In each panel, the Z01 evolutionary tracks are plotted as a solid black line. Where  predictions for the used filters were available we included the M08 tracks (dashed lines) as well. Ages are labelled along the tracks. The black dots are clusters with regular SEDs (from UV to IR). The red triangles are cluster with an excess at $\lambda >1.0$ $\mu$m. The blue diamonds shows objects with an excess starting longword $\lambda \geq 0.8$ $\mu$m. The black arrows indicate an visual extinction of A$_V=1$. Mean errors are included. A solid red line divides the plots in two regions: $R-I>0$ (older than 10 Myr) and  $R-I<0$ (younger than 10 Myr). }
\end{figure}

A significant fraction of young star clusters in Haro\,11 \citep{2010MNRAS.407..870A}, ESO\,185-IG13 \citep{2011MNRAS.414.1793A}, and Mrk\,930 \citep{2011MNRAS.tmp..739A} shows a clear signature of a flux excess at near-IR wavebands. The models used \citep[][]{2001A&A...375..814Z, 2010ApJ...725.1620A} include a self-consistent treatment of the the photoionized gas, important during the first few Myrs of the cluster evolution. However, these models are not able to reproduce the NIR observed fluxes of the clusters. In Fig.~\ref{fig1}, we show two representative cases of cluster spectral energy distributions (SEDs). A large fraction of analysed clusters in the 3 galaxies have regular SEDs, easily fitted by our models (see left panel,Fig.~\ref{fig1}). However, a considerable number of clusters have SEDs similar to cluster \# 43 in the right panel of Fig.~\ref{fig1}. Three different sets of fits have been performed for each cluster \citep[see][for details]{2010MNRAS.407..870A, 2011MNRAS.414.1793A} and depending on the outcomes, clusters have been divided into 3 samples: 1) regular SEDs have been fitted from UV to IR; 2) if the source presented an excess at $\lambda > 1.0$ $\mu$m, the fit was performed excluding the IR data (so called IR  excess); 3) if the excess affected also $I$ band ($\sim 8000$ \AA), then only the data from UV to optical ($\lambda < 8000$ \AA) were included in the fit (referred to as NIR excess). The exclusion of the data points with an excess was necessary in order to reproduce the observed UV and optical SEDs of the clusters. The two observed fluxes at 0.5 and 0.6 $\mu$m of cluster \# 43 are produced, respectively, by a narrow filter (F550M) centred on the continuum and a filter (F606W) which transmits H$\alpha$. A jump between these two filters can be used as an age indicator (only clusters younger than 10 Myr have H$\alpha$ in emission). Clearly,  \# 43 is a very young cluster. However, if the NIR data points (including $I$ band) are included in the fit, the age obtained for this cluster is 45 Myr \citep[see][for details]{2010MNRAS.407..870A}.  In Table~\ref{tab1}, we give the fractions of clusters with a regular SED or NIR excess.
\begin{table}[ht] 
\caption{Fraction of clusters in each SED sample, of the three galaxies.} 
\center
\begin{minipage}{0.5\textwidth}
\center
\begin{tabular}{cccc} 
\hline\hline 
targets & regular SEDs & IR excess & NIR excess\\   
& \%& \%& \%\\
\hline 
Haro\,11&44&14&42\\
ESO\,185&68&10&22\\
Mrk\,930&62&12&26\\
\hline
\end{tabular} 
\end{minipage}
\label{tab1} 
\end{table}
Color-color diagrams of the whole Mrk\,930 cluster population \citep[see][for s complete analysis]{2011MNRAS.tmp..739A} are shown in Fig.~\ref{fig2}.  The $R-I$ color is used as reference. Generally, a negative $R-I < 0$ color indicates ages younger than 10 Myr and is produced by a strong nebular contribution to the $R$ filter, which includes the bright H$\alpha$ line. However, this assumption is not always valid due to the flux excess in the $I$ band. The clusters, in the diagrams, have different symbols depending of their observed SEDs: black dots indicate normal SEDs, red triangles clusters with an excess at $\lambda \geq 1.0$ $\mu$m (hereafter, IR excess), and blue diamonds SEDs which deviate at $\lambda \geq 0.8$ $\mu$m (hereafter, NIR excess). We notice that clusters with a NIR excess have a $R-I$ color between 0.2 and 1.2 mag redder than the prediction made by the best-fitting SED model. The $R-I$ color of the "blue" cluster-diamonds is such that, if overlooked, causes age (and mass) overestimates, affecting the results of the optical based cluster analysis.   The $UV-R$ (left panel) color shows that the clusters detected in the $FUV$ are young, at least younger than 30-40 Myr. Many of the clusters with an NIR excess are located in an area where $R-I > 0$, e.g., corresponding to ages older than 10 Myr. However, their UV color is compatible with being a few Myrs old.  The $FUV$ band is sensitive to the reddening. Looking at the color-color diagram, one can see that clusters detected in the $FUV$ have extinctions A$_V \leq 1.0$ ( e.g., the arrow in the plot). Therefore, we consider these very young $FUV$ bright clusters as systems which have already gone through the deeply embedded phase. The inclusion of the IR color (right panel), clearly shows that clusters with a flux excess (clearly young) at the redder wavelengths are mainly located in an area of the diagram where the $R-I > 0$ and $H-R > 1.0$ mag, e.g. apparently older than 1 Gyr. The IR color of the clusters with a red excess suggests that the extinction in these objects should be much higher than the one predicted by the optical and UV colors. In other words, at the NIR wavelengths it is possible that we are probing a different, more deeply embedded stellar populations, indicating that these are very young clusters.

\section{Possible origin of the NIR excess}

\subsection{The $I$ band excess} The $I$ band excess has been found only in very young clusters (usually $<6$ Myr). A viable explanation for this feature is the extended red emission (ERE, see for a review \citealp{2004ASPC..309..115W}).  The ERE is observed as a soft rising continuum between $0.6-0.9$ $\mu$m. It is observed around  galactic and extragalactic H{\sc ii} regions and caused by a photoluminescence reaction on dust grains heated by hard UV radiation. Such energetic photons are mainly produced in short-lived massive stars. This could explain why the $I$ band excess in our clusters is over after 6 Myr. 

\subsection{The IR excess} Several mechanisms can concur to make the flux at $\lambda > 1.0$ $\mu$m higher than predicted by models. The distance of the galaxy and the resolution achieved - the best with the current accessible facilities - limit our studies to the integrated properties of these YSCs. However, observations of close-by resolved clusters and numerical predictions of stellar populations in clusters can give us an hint of the mechanisms which are likely producing the observed excess.

Among the youngest and massive resolved star clusters, 30 Doradus (hereafter 30 Dor) in the Large Magellanic Cloud (LMC), represents the best reference to understand what a recently born star cluster looks like. 30 Dor is the central region of the extended Tarantula nebula. Multiwavelength studies of this regions have dissected the different components of the complex 30 Dor environment. \citet{1997ApJS..112..457W} identified five different stellar populations in the region: the bright core early O-type stars which are part of the compact star cluster R136; in the north and west region embedded massive YSOs; 3 more evolved stellar population groups in the southern and 1.0' away in the western region. The R136 cluster has a mass of $\sim 10^5$ M$_{\odot}$ and is 3 Myr old. This nuclear region ($\leq 3$ pc) is dust and gas-free. However, the aperture radius we are using in our analysis is much larger (radius of $\approx 36$ pc) than the size of R136. Our apertures are comparable to the size of the image of 30 Dor showed in Figure 1 of \citet{2010MNRAS.405..421C}. This suggests that while the optical range is dominated by the stellar and gas emission, the IR transmits also flux from the diffuse dust, heated by the hard UV radiation field, the embedded YSOs formed in triggered star formation events at the edge of the nucleus, where most of the dense gas and dust filaments are located, and the low mass stars still in the PMS phase. 

In the literature, studied cases of IR excess in young embedded or partially embedded unresolved extragalactic clusters have explained the red excess as due to an important contribution by YSOs \citep{2009MNRAS.392L..16F}, or hot dust (\citealp{2005A&A...433..447C}). Likely, the same mechanism is causing the excess in young star clusters in BCGs.

After several Myr this complex phase is over, so it cannot explain why we  still observe objects with an IR excess at older ages. For these evolved clusters a possible source of excess can be an important contribution from red super giants (RSGs). Models usually predict the number of RSGs, assuming that the stars in a cluster fully populate the underlying initial mass function (IMF). However, this assumption is not valid, if the cluster is less massive than $10^4$ M$_{\odot}$, and causes important variations for cluster masses below $5\times10^3$ M$_{\odot}$ \citep{2011A&A...529A..25S}. Moreover, it has been observed that in metal-poor environments the number of observed RSGs tends to be higher than the predicted one from the ratio of blue versus RSGs. Therefore, both effects would be observed as a rise in the NIR integrated flux of an unresolved cluster, which our current models cannot fully account for. It is not clear, however, why we do not see any mass dependence between the excess in $H$ band and the mass of the cluster, which could support the stochasticity scenario, or why our models cannot predict the correct number of RSGs only for some of the clusters. NIR spectroscopy is needed to test these scenarios.

Another possible explanation is that a second stellar population is forming in the clusters. If a second population is forming now, then we expect to detect it mainly in the NIR. Multiple stellar populations have been detected in globular clusters and, recently, even in young star clusters \citep{2011arXiv1106.4560L}. Are we observing the formation of a new stellar population in these massive YSCs? 

Finally, a bottom heavy IMF, i.e. a much steeper slope at low mass ranges (higher number of low mass stars), could also explain the excess. However, studies of IMF variations in massive star clusters (see \citealp{2010ARA&A..48..339B}) have not found any confirmed case of such type of cluster. Moreover, it is not clear how, in the same galaxy, a fraction of clusters could form with a different IMF.

%
%
\small  
%
%

%
%
%
%
%

\bibliographystyle{aa}
\bibliography{mnemonic,ref_user}

\begin{thebibliography}{20}
\expandafter\ifx\csname natexlab\endcsname\relax\def\natexlab#1{#1}\fi

\bibitem[{{Adamo} {et~al.}(2011{\natexlab{a}}){Adamo}, {{\"O}stlin},
  {Zackrisson}, \& {Hayes}}]{2011MNRAS.414.1793A}
{Adamo}, A., {{\"O}stlin}, G., {Zackrisson}, E., \& {Hayes}, M.
  2011{\natexlab{a}}, \mnras, 414, 1793

\bibitem[{{Adamo} {et~al.}(2010{\natexlab{a}}){Adamo}, {{\"O}stlin},
  {Zackrisson}, {Hayes}, {Cumming}, \& {Micheva}}]{2010MNRAS.407..870A}
{Adamo}, A., {{\"O}stlin}, G., {Zackrisson}, E., {et~al.} 2010{\natexlab{a}},
  \mnras, 407, 870

\bibitem[{{Adamo} {et~al.}(2011{\natexlab{b}}){Adamo}, {{\"O}stlin},
  {Zackrisson}, {Papaderos}, {Bergvall}, {Rich}, \&
  {Micheva}}]{2011MNRAS.tmp..739A}
{Adamo}, A., {{\"O}stlin}, G., {Zackrisson}, E., {et~al.} 2011{\natexlab{b}},
  \mnras, 739

\bibitem[{{Adamo} {et~al.}(2010{\natexlab{b}}){Adamo}, {Zackrisson},
  {{\"O}stlin}, \& {Hayes}}]{2010ApJ...725.1620A}
{Adamo}, A., {Zackrisson}, E., {{\"O}stlin}, G., \& {Hayes}, M.
  2010{\natexlab{b}}, \apj, 725, 1620

\bibitem[{{Bastian} {et~al.}(2010){Bastian}, {Covey}, \&
  {Meyer}}]{2010ARA&A..48..339B}
{Bastian}, N., {Covey}, K.~R., \& {Meyer}, M.~R. 2010, \araa, 48, 339

\bibitem[{{Bik} {et~al.}(2010){Bik}, {Puga}, {Waters}, {Horrobin}, {Henning},
  {Vasyunina}, {Beuther}, {Linz}, {Kaper}, {van den Ancker}, {Lenorzer},
  {Churchwell}, {Kurtz}, {Kouwenhoven}, {Stolte}, {de Koter}, {Thi},
  {Comer{\'o}n}, \& {Waelkens}}]{2010ApJ...713..883B}
{Bik}, A., {Puga}, E., {Waters}, L.~B.~F.~M., {et~al.} 2010, \apj, 713, 883

\bibitem[{{Bressert} {et~al.}(2010){Bressert}, {Bastian}, {Gutermuth},
  {Megeath}, {Allen}, {Evans}, {Rebull}, {Hatchell}, {Johnstone}, {Bourke},
  {Cieza}, {Harvey}, {Merin}, {Ray}, \& {Tothill}}]{2010MNRAS.409L..54B}
{Bressert}, E., {Bastian}, N., {Gutermuth}, R., {et~al.} 2010, \mnras, 409, L54

\bibitem[{{Campbell} {et~al.}(2010){Campbell}, {Evans}, {Mackey}, {Gieles},
  {Alves}, {Ascenso}, {Bastian}, \& {Longmore}}]{2010MNRAS.405..421C}
{Campbell}, M.~A., {Evans}, C.~J., {Mackey}, A.~D., {et~al.} 2010, \mnras, 405,
  421

\bibitem[{{Carlson} {et~al.}(2007){Carlson}, {Sabbi}, {Sirianni}, {Hora},
  {Nota}, {Meixner}, {Gallagher}, {Oey}, {Pasquali}, {Smith}, {Tosi}, \&
  {Walterbos}}]{2007ApJ...665L.109C}
{Carlson}, L.~R., {Sabbi}, E., {Sirianni}, M., {et~al.} 2007, \apjl, 665, L109

\bibitem[{{Cresci} {et~al.}(2005){Cresci}, {Vanzi}, \&
  {Sauvage}}]{2005A&A...433..447C}
{Cresci}, G., {Vanzi}, L., \& {Sauvage}, M. 2005, \aap, 433, 447

\bibitem[{{Elmegreen}(2010)}]{2010IAUS..266....3E}
{Elmegreen}, B.~G. 2010, in IAU Symposium, Vol. 266, IAU Symposium, ed. {R.~de
  Grijs \& J.~R.~D.~L{\'e}pine}, 3--13

\bibitem[{{Elmegreen}(2011)}]{2011arXiv1101.3112E}
{Elmegreen}, B.~G. 2011, ArXiv e-prints

\bibitem[{{Fern{\'a}ndez-Ontiveros} {et~al.}(2009){Fern{\'a}ndez-Ontiveros},
  {Prieto}, \& {Acosta-Pulido}}]{2009MNRAS.392L..16F}
{Fern{\'a}ndez-Ontiveros}, J.~A., {Prieto}, M.~A., \& {Acosta-Pulido}, J.~A.
  2009, \mnras, 392, L16

\bibitem[{{Gieles} \& {Portegies Zwart}(2011)}]{2011MNRAS.410L...6G}
{Gieles}, M. \& {Portegies Zwart}, S.~F. 2011, \mnras, 410, L6

\bibitem[{{Larsen} {et~al.}(2011){Larsen}, {de Mink}, {Eldridge}, {Langer},
  {Bastian}, {Seth}, {Smith}, {Brodie}, \& {Efremov}}]{2011arXiv1106.4560L}
{Larsen}, S.~S., {de Mink}, S.~E., {Eldridge}, J.~J., {et~al.} 2011, ArXiv
  e-prints

\bibitem[{{{\"O}stlin} {et~al.}(2003){{\"O}stlin}, {Zackrisson}, {Bergvall}, \&
  {R{\"o}nnback}}]{2003A&A...408..887O}
{{\"O}stlin}, G., {Zackrisson}, E., {Bergvall}, N., \& {R{\"o}nnback}, J. 2003,
  \aap, 408, 887

\bibitem[{{Silva-Villa} \& {Larsen}(2011)}]{2011A&A...529A..25S}
{Silva-Villa}, E. \& {Larsen}, S.~S. 2011, \aap, 529, A25+

\bibitem[{{Walborn} \& {Blades}(1997)}]{1997ApJS..112..457W}
{Walborn}, N.~R. \& {Blades}, J.~C. 1997, ApJS, 112, 457

\bibitem[{{Witt} \& {Vijh}(2004)}]{2004ASPC..309..115W}
{Witt}, A.~N. \& {Vijh}, U.~P. 2004, in Astronomical Society of the Pacific
  Conference Series, Vol. 309, Astrophysics of Dust, ed. {A.~N.~Witt,
  G.~C.~Clayton, \& B.~T.~Draine}, 115--+

\bibitem[{{Zackrisson} {et~al.}(2001){Zackrisson}, {Bergvall}, {Olofsson}, \&
  {Siebert}}]{2001A&A...375..814Z}
{Zackrisson}, E., {Bergvall}, N., {Olofsson}, K., \& {Siebert}, A. 2001, \aap,
  375, 814

\end{thebibliography}

\end{document}